\documentclass[a4paper,fleqn,usenatbib,useAMS]{mnras}
\usepackage[T1]{fontenc}
\usepackage{ae,aecompl}
\usepackage{graphicx}	
\usepackage{amsmath}	
\usepackage{amssymb}
\usepackage{newtxtext,newtxmath}
\usepackage{multicol}
\usepackage{bm}
\usepackage{pdflscape}

\newcommand{\be}{\begin{equation}}
\newcommand{\ba}{\begin{eqnarray}}
\newcommand{\ee}{\end{equation}}
\newcommand{\ea}{\end{eqnarray}}

\newcommand{\gtsima}{$\; \buildrel > \over \sim \;$}
\newcommand{\ltsima}{$\; \buildrel < \over \sim \;$}
\newcommand{\gsim}{\lower.5ex\hbox{\gtsima}}
\newcommand{\lsim}{\lower.5ex\hbox{\ltsima}}
\newcommand{\Lya}{Ly~$\alpha$}
\newcommand{\msun}{\rm{M}_{\sun}}

\begin{document}

\title[Signatures of reionization in the LG]{Reionization of the Milky Way, M31, and their satellites I: reionization history and star formation}
\author[K.~L.~Dixon, et al.]
{Keri L. Dixon$^{1}$\thanks{e-mail: K.Dixon@sussex.ac.uk},
Ilian T. Iliev$^{1}$,
Stefan Gottl\"ober$^{2}$,
Gustavo Yepes$^{3,4}$,
\newauthor Alexander Knebe$^{3,4}$,
Noam Libeskind$^{2}$, and
Yehuda Hoffman$^{5}$
\\
$^1$ Astronomy Centre, Department of Physics \& Astronomy, Pevensey II Building, University of Sussex, Falmer, Brighton, BN1 9QH, \\UK
\\
$^2$Leibniz-Institut f\"{u}r Astrophysik, 14482 Potsdam, Germany\\
$^3$Departamento de F\'isica Te\'{o}rica, M\'{o}dulo 15, Facultad de Ciencias, Universidad Aut\'{o}noma de Madrid, 28049 Madrid, Spain \\
$^4$Astro-UAM, UAM, Unidad Asociada CSIC \\
$^5$Racah Institute of Physics, Hebrew University, 91904 Jerusalem, Israel \\
}
\twocolumn
\date{\today}
\pubyear{2016} \volume{000} \pagerange{1}

\voffset-.6in
\maketitle\label{firstpage}

\begin{abstract}
Observations of the Milky Way (MW), M31, and their vicinity, known as the Local Group (LG), can provide clues about the sources of reionization. We present a suite of radiative transfer simulations based on initial conditions provided by the Constrained Local UniversE Simulations (CLUES) project that are designed to recreate the Local Universe, including a realistic MW-M31 pair and a nearby Virgo. Our box size (91~Mpc) is large enough to incorporate the relevant sources of ionizing photons for the LG. We employ a range of source models, mimicking the potential effects of radiative feedback for dark matter haloes between $\sim\!10^8-10^9\,\msun$. Although the LG mostly reionizes in an inside-out fashion, the final 40 per cent of its ionization shows some outside influence. For the LG satellites, we find no evidence that their redshift of reionization is related to the present-day mass of the satellite or the distance from the central galaxy. We find that less than 20 per cent of present-day satellites for MW and M31 have undergone any star formation prior to the end of global reionization. Approximately five per cent of these satellites could be classified as fossils, meaning the majority of star formation occurred at these early times. The more massive satellites have more cumulative star formation prior to the end of global reionization, but the scatter is significant, especially at the low-mass end. Present-day mass and distance from the central galaxy are poor predictors for the presence of ancient stellar populations in satellite galaxies.\end{abstract}


\section{Introduction}
\label{sec:intro}

Hydrogen reionization, a major phase transition for the Universe, induced significant changes in the intergalactic medium (IGM). This process is highly spatially inhomogeneous, with the local reionization history correlated with the environment of a given region. The local ionization state impacts the local temperature and is related the ionizing flux from both near and far sources. Therefore, the timing and morphology of reionization impacts galaxy formation and evolution. Clues from recent indirect measures have suggested that the epoch of reionization is likely extended in time, with the bulk occurring in the range $6\lesssim z \lesssim 10$ \citep[e.g.][]{Robe15a, Bouw15a, Mitr15a}. These observations include high-redshift quasar spectra \citep[e.g.][]{Fan06, Mort11a, Bolt11a, McGr15a}, the cosmic microwave background (CMB) polarization \citep{Koma11a,Plan16a}, IGM temperature measurements \citep[e.g.][]{Theu02, Rask12a, Bolt12a}, and the decline of Lyman~$\alpha$ (\Lya) emission in high-redshift galaxies \citep[e.g.][]{Star10a, Sche12a, Pent14a, Tilv14a}. 

The process of reionization particularly affects low-mass objects, whose shallow gravitational potentials make their baryon content more susceptible to outside influence. Specifically, as reionization progresses, the number of ionizing photons increases, and the IGM is heated, which locally raises the Jeans mass. As has been extensively studied, this increase delays and possibly suppresses the ability of low-mass haloes to form stars (in some cases permanently, via photoevaporation), regulating ongoing reionization \citep[e.g.][]{Efst92a, Bark99a, Gned00a, Shap04a, Ilie05b}. The exact threshold below which star formation cannot proceed and nature of this effect -- a sharp or gradual cutoff, full photoevaporation, or a shielded inner region, etc. -- is debated \citep[e.g.][]{Susa04a, Hoef06a,Okam08a, Xu16a,Finl17a}. Consequently, the observational signatures of this suppression are unclear. Furthermore, as reionization is inhomogeneous and extended, the exact impact of reionization on a specific dwarf galaxy is environmentally dependent and very difficult to assess.

The Local Group (LG), referring to the Milky Way (MW) and M31 and their immediate vicinity, encompasses dwarf satellite galaxies that are within the range of masses that should be affected by reionization. Furthermore, their close proximity to us makes measurement of their star formation histories (SFHs) possible. Generally, no widespread signature of reionization, strictly interpreted as a cutoff for star formation around the time of reionization ($\lesssim\!13$~Gyr ago), is observed \citep{Greb04a, Apar16a}; though many dwarf galaxies exhibit a degree of quenching that is potentially consistent with reionization \citep{Weis14b, Brow14a, Skil16a}. The exact nature of the effect of reionization on LG dwarf galaxies has been studied extensively from a theoretical perspective with predictions ranging from full to partial quenching of star formation \citep[e.g.][]{Rico05a, Bovi11a, Simp13a, Beni15a}.

For the LG as a whole, one important question is whether it is reionized from the inside out or from the outside in. Past studies have generally found inside-out reionization by the LG's own sources, though exact assumptions about the efficiency of ionizing sources can alter this conclusion \citep{Ilie11a,Ocvi14a}. Beyond purely theoretical considerations, this behaviour has important consequences for galaxy formation simulations, including the local ionizing flux and timing of reionization. Since reionization is inhomogeneous and environmentally dependent, the most realistic distribution of nearby (especially large) objects is necessary for an accurate picture of the history of our LG. Many simulations have a small volume and/or only include one halo or an MW-M31 pair of haloes \citep{Gned06a,Okam09a,Bovi11b,Ocvi13a}. These simulations, even if they include the effects of inhomogeneous reionization, may miss the larger characteristic scales of this process \citep{Ilie14a}. 

Many of the theoretical studies of the LG and associated satellites and dwarf galaxies use semi-analytic methods \citep[e.g.][]{Bull00a, Bens02a, Some02a} or an $N$-body simulation with simplistic reionization \citep[e.g.][]{Krav04a, Mada08a}. Recent numerical studies have included baryonic effects, finding good agreement with observed satellites \citep[e.g.][]{Sawa10a, Sawa16a,Wetz16a}. These all fail, however, to include realistic, inhomogeneous reionization. To study the question of the reionization of the LG and the satellites therein, we employ a dark matter (DM) simulation, produced by the CLUES\footnote{Constrained Local UniversE Simulations: \url{www.clues-project.org}} collaboration, that is large enough to reproduce the mean global reionization history, nearly $100^3$~Mpc$^3$ in volume \citep{Ilie14a}. Specifically, the initial conditions are constrained to reproduce the spatial and velocity structure of the LG and its vicinity at the present day. Importantly, a realistic MW, M31, Virgo (the largest nearby cluster), local void, and Virgo filament are present, so their mutual influences are captured. Even though satellites may not be within the virial radius of the MW or M31 during reionization \citep{Wetz15a}, the character of the neighbourhood of the LG progenitors affects the reionization history of these satellites. 

We combine this DM realization with full radiative transfer (RT) simulations to track the evolution of the LG, including the MW and M31 and their satellites, throughout cosmic reionization. In this work, we are interested in what imprints the radiative feedback on low-mass galaxies might have left. We, therefore, separate the ionizing sources into two distinct populations: high-mass ones, with DM halo masses above $\sim10^9\,\msun$ (high-mass, atomic-cooling haloes, or `HMACHs') that are unaffected by radiative feedback and those between approximately $10^8$ and $10^9\,\msun$ (low-mass, atomic-cooling haloes, or `LMACHs') that are susceptible to photoionizing radiation. The $10^8\,\msun$ mass limit roughly corresponds to a virial temperature of $10^4$\,K, below which the halo gas is unable to radiatively cool through hydrogen and helium atomic lines. Throughout this work, we explore four distinct source models, varying the efficiencies and degree of star formation suppression from reionization.

Recently, \citet{Ocvi16a} presented the Cosmic Dawn simulation, the first fully coupled RT and hydrodynamics simulation of the Local Universe. The same constrained initial conditions as our simulations presented here were used, and the Cosmic Dawn results are complementary. The radiative and supernovae feedback in Cosmic Dawn is treated dynamically and is, thus, more realistic and detailed. However, due to the large computational expense of fully coupled simulations, Cosmic Dawn is just a single simulation, using a particular set of physics parameters; while our simulations investigate the effects of different subgrid models for the ionizing sources and their suppression. The underlying $N$-body simulation here also has significantly better spatial resolution than Cosmic Dawn (which employs a particle-mesh gravity on a fixed grid), allowing us to more reliably identify all star-forming satellite galaxy progenitors.

The outline of the paper is as follows. In Section~\ref{sec:method}, we outline in detail our numerical methods, including our source models. We present our global, as in the entire box, results in  Section~\ref{sec:global}. Section~\ref{sec:local} details the reionization history and star formation for satellites during reionization for the LG. In Section~\ref{sec:obs}, we compare our predictions with  observational data. We then conclude in Section~\ref{sec:summary}.

\section{Methodology}
\label{sec:method}

In this section, we outline the numerical methods used in this paper. First, we describe the underlying $N$-body simulations created to match the Local Universe at $z=0$. Second, we introduce our RT methods and the use of the $N$-body data products. Last, we describe our source models in detail, specifically our radiative feedback assumptions.

\subsection{Constrained simulations of the Local Universe}
\label{sec:CLUES}

\begin{table*}
\caption{The present-day MW and M31 results.}
\label{tab:MW_M31}
\begin{center}
\begin{tabular}{@{}llllllllll}\hline
\hline
galaxy & $M_{z=0}$ & x & y & z &  $N_{z=0}^{\rm sats}$ & $N_{z=6.5}^{\rm sats}$ \\
& \tiny{($\msun$)} &  & \tiny{(Mpc)} & && &
\\[1.5mm]
\hline
MW & $1.13\times10^{12}$ & 31.14 & 46.73 & 46.84 & 120 & 23
\\[2mm]
M31 & $1.59\times10^{12}$ & 30.90 & 46.29 & 47.41 & 151 & 23
\\%[2mm] 
\hline
\end{tabular}
\end{center}
\begin{flushleft}
\end{flushleft}
\end{table*}

\begin{table*}
\caption{Reionization simulation parameters and global reionization history results.}
\label{tab:summary}
\begin{center}
\begin{tabular}{@{}llllllllllllll}\hline
\hline
label & run & box size & $g_{\gamma}$ & $g_{\gamma}$ & $g_{\gamma}$ &mesh  &  $\tau_{\rm es}$ & $z_{10\%}$&$z_{50\%}$&$z_{90\%}$&$z_{\rm reion}$ \\
&      & \tiny{($h^{-1}$Mpc)}  & \tiny{HMACH} & \tiny{LMACH} & \tiny{LMACH$_{\rm supp}$} & &  & &&&
\\[1.5mm]
\hline
LG1 & 91Mpc\_g1.7\_0 & 64 & 1.7 & 0 & 0 & $256^3$ & 0.052 & 7.909 & 6.793 & 6.354 & 6.172
\\[2mm]
LG2 & 91Mpc\_g1.7\_7.1S & 64 & 1.7 & 7.1 & 0 & $256^3$ & 0.053 & 8.172 & 6.905 & 6.418 & 6.231
\\[2mm] 
LG3 & 91Mpc\_g1.7\_7.1pS & 64 & 1.7 & 7.1 & 1.7 & $256^3$ & 0.062 & 9.026 & 7.712 & 7.180 & 7.020
\\[2mm]
LG4 & 91Mpc\_g1.7\_gS & 64 & 1.7 & 1.7 & equation~(\ref{eq:grad_eff}) & $256^3$ & 0.055 & 8.340 & 7.139 & 6.651 & 6.483
\\%[2mm] 
\hline
\end{tabular}
\end{center}
\begin{flushleft}
\end{flushleft}
\end{table*}

The dark matter density and halo fields used in this work were extracted from a constrained simulation performed within the CLUES project \citep{Gott10a,Yepe14a}. The background cosmology is $\Omega_{\rm m} = 0.279, h=0.7, \Omega_{\rm b}=0.046, \sigma_8 =0.817, n=0.96$.  The initial conditions were set to reproduce an MW and M31 galaxy pair, as well as the Virgo cluster, consistent with observations. Constrained simulations such as these use the observed radial velocities of nearby galaxies as constraints for the generation of initial conditions. These data only constrain scales larger than a few Mpc. Therefore, structures such as the Virgo cluster are well reproduced; while smaller structures, such as the LG itself, are  unconstrained. The procedure to obtain a `realistic' LG embedded in the correct constrained environment proceeds as follows (for details see \citealt{Yepe14a}). In a low-resolution, constrained simulation, first the Virgo cluster is identified, and then, we search for an object which closely resembles the LG. The selected simulation is then repeated with higher resolution. 

With these constrained initial conditions, we evolve with \textsc{\small GADGET-3} \citep{Spri05a} a cosmological box of 91~Mpc on a side with $2048^3$ particles that have a particle mass of $3.37\times10^6\,\msun$. In total 209, simulation outputs were stored between $z=31$ and $z=0$. The same initial conditions, but in different mass and spatial resolutions, were used for the simulations discussed in \citet{Ocvi16a}.

In the next step, we identify DM halos in all outputs. We use the \textsc{\small AHF} halo finder, based on spherical overdensity method \citep{Knol09a}. At $z\geqslant6$, DM haloes more massive than $1.43\times10^8\,\msun$ that will host reionization sources have been taken into account for the subsequent RT simulations. For the RT calculations, the DM haloes and density fields are smoothed to a 256$^3$ grid.

The resulting properties of the simulated AHF positions and masses of the MW and M31 are summarized in Table~\ref{tab:MW_M31}. The majority of measurements of the MW and M31 suggest that M31 is slightly more massive, see Table A1 and A2 of \citet{Carl17a}. Therefore, we identify the more massive halo as M31 and the less massive halo as the MW.

\subsection{RT methods and source models}
\label{sec:sources}

Given the underlying DM fields and haloes as described in section~\ref{sec:CLUES}, we apply RT to track the reionization history of all cells using four different source models, starting at $z=20.586$ and ending when the global ionization fraction exceeds 99 per cent. The RT simulations are performed with our code \textsc{\small C$^2$-Ray} \citep[Conservative Causal Ray-Tracing;][]{Mell06a}. The code has been tested in detail against a number of exact analytical solutions \citep{Mell06a}, as well as in direct comparison with a number of other independent RT methods on a standardised set of benchmark problems \citep{Ilie06b,Ilie09a}. 

Table~\ref{tab:summary} summarizes the four models for the ionizing sources, previously discussed in detail in \citet{Dixo16a}. Here, HMACHs are defined to be all haloes above $1.43\times10^9\,\msun$, and LMACHs are all haloes below this and above our minimum threshold from the previous section ($1.43\times10^8\,\msun$). All identified DM haloes are potential sources of ionizing radiation. We assume that the source emissivities are proportional to the host halo mass with an effective (potentially mass-dependent) mass-to-light ratio, with different values adopted for LMACHs and HMACHs. These efficiencies are chosen to conclude reionization before $z = 6$. For all haloes in the simulation volume, each halo that is not suppressed by Jeans-mass filtering is an ionizing source. For a source with a DM halo mass, $M_{\rm halo}$, we assign ionizing photon emissivity, $\dot{N}_\gamma$, according to
\be
\dot{N}_\gamma=g_\gamma\frac{M_{\rm halo}\Omega_{\rm b}}{m_{\rm p}(10\,\rm Myr)\Omega_0},
\label{eq:Ndot}
\ee
where $m_{\rm p}$ is the proton mass. The proportionality coefficient, $g_{\gamma}$, reflects the ionizing photon production efficiency of the stars per stellar atom, $N_{\rm i}$, the star formation efficiency, $f_*$, the source lifetime, $t_{\rm s}$, and the escape fraction, $f_{\rm esc}$ \citep[e.g.][]{Haim03a,Ilie12a}:
\be
g_\gamma=f_*f_{\rm esc}N_{\rm i}\left(\frac{10 \;\mathrm{Myr}}{t_{\rm s}}\right)
\ee
\label{eq:eff}
and can additionally include mass dependence. In our simulations, $t_{\rm s} = 11.53$~Myr, or equivalent to one timestep. We set $g_\gamma$ based on the source models described below, and the exact values for the remainder of the variables in the right-hand side of equation~(\ref{eq:eff}) are unimportant from a RT standpoint. We will return to these quantities in section~\ref{sec:SF}.

Our full simulation notation reads $Lbox\_gI\_(J)(Supp)$ (the bracketed quantities are listed only when needed), where $'Lbox'$ is the simulation box size in comoving Mpc, $'I'$ and $'J'$ are the values of the $g_{\gamma}$ factor for HMACHs and LMACHs, respectively. The symbol $`Supp'$ indicates the suppression model, the details of which are described below. Throughout the paper, we refer to the simulations using the short-hand in the left column of Table~\ref{tab:summary}. 

Each source model imposes different suppression criteria and ionizing photon production efficiencies, as follows:
\begin{itemize}
\item{\raggedright{}\textit{HMACHs only:}}

Here, only HMACHs produce ionizing photons, corresponding to reionization being driven exclusively by relatively large galaxies. In terms of source suppression, this model could be considered an extreme case where all LMACHs are fully suppressed (or never formed) at all times. Physically, this model could arise when mechanical feedback from supernovae quenches the star formation in low-mass haloes quickly. All HMACHs have a source efficiency of $g_\gamma = 1.7$. For notation, $J=0$ in this case.
\\
 
\item{\raggedright{}\textit{Fully suppressed LMACHs (S):}}

HMACHs are once again assigned $g_\gamma = 1.7$. LMACHs are assigned a higher efficiency $g_\gamma = 7.1$ in neutral regions to mimic the properties of early galaxies that likely have higher photon production efficiencies overall, arising from massive, Pop~III stars and/or higher escape fractions\footnote{Note that later in the paper we assume a constant $f_{\rm esc}$ for all haloes. For overall progression of reionization, the distinction does not matter, and many of our conclusions would remain unchanged.}. We assume that LMACHs in ionized regions produce no ionizing photons, corresponding to the case of aggressive suppression of low-mass haloes from either mechanical or radiative feedback or a combination thereof.
\\

\item{\raggedright{}\textit{Partially suppressed LMACHs (pS):}} 

For this model, LMACHs are assumed to contribute to reionization at all times. In neutral regions, we assign LMACHs a higher efficiency as in the previous model, $g_\gamma = 7.1$. In ionized regions, these small galaxies are suppressed, resulting in diminished efficiency, and we set this efficiency to the same as the HMACHs, $g_\gamma = 1.7$. One physical situation that may be represented by this model is that the fresh gas supply is diminished or cut off by the photoheating of surrounding gas, but a gas reservoir within the galaxy itself remains available for star formation.
\\

\item{\raggedright{}\textit{Mass-dependent suppression of LMACHs (gS):}}

Instead of a sharp decrease in ionizing efficiency as in the previous two cases, we also consider the gradual, mass-dependent suppression of sources in ionized regions. As before, HMACHs are assigned $g_\gamma = 1.7$ everywhere, and LMACHs have that same efficiency when residing in neutral regions. In ionized patches, LMACHs are suppressed in a mass-dependent manner loosely following \citet{Wise09a} and \citet{Sull16a} *et. al, will change bib at end* (in prep.):
\be
 g_\gamma = g_{\gamma,\rm HMACH} \times \left[\frac{M_{\rm halo}}{9\times10^8\,h^{-1}\msun}-\frac{1}{9}\right],
 \label{eq:grad_eff}
\ee
essentially linear in logarithmic units of halo mass with $g_\gamma = g_{\gamma,\rm HMACH}$ at $10^9\,\msun$ and $g_\gamma = 0$ at $10^8\,\msun$.

\end{itemize}

\subsection{Satellites at \textit{z}~=~0 and high-redshift progenitors}
\label{sec:sats}

After completing the RT simulations, we are interested in the reionization histories of the satellites of the MW and M31. We use the hierarchical friends-of-friends (hFoF) algorithm (Appendix B of \citealt{Rieb13a}) to identify the satellites in DM haloes at redshift zero and their progenitors at higher redshift. The hFoF algorithm has been designed to find substructures of FoF halos at higher overdensities and performs well locating substructure as compared to other methods \citep{Kneb11a}. To this end, the FoF algorithm is performed with a series of different linking lengths, which identify objects at higher overdensities. The advantage of this procedure is that these high density FoF objects are defined in a unique way, independent of the position of the satellite inside or outside the hosting FoF halo. Due to the FoF algorithm, any particle belongs to one FoF object only, which allows a direct construction of the progenitor-successor relations for FoF objects. 

Here, we use a linking length of 0.025 to identify all substructures in the DM haloes and up to  0.3~Mpc of the $z=0$ positions of the MW or M31. These substructures have 512 times higher overdensities than the hosting haloes, defined with the standard linking length of 0.2. Thus, their mass is substantially smaller than a FoF mass from the larger linking length. Note that 0.3~Mpc was chosen to match the convention of \citet{McCo12a}. At $z = 0$, we find 120 and 151 satellites for the MW and M31, respectively. We have assumed a minimum of 20 particles for an identified FoF object. 

As progenitors of the satellites, we consider FoF objects defined with a linking length of 0.2 at high redshifts. These objects have approximately virial overdensity at these redshifts and are very similar in position and mass to the AHF objects that we used in our RT simulations. We include all progenitors above $1.43\times10^8\msun$ for each $z=0$ object.

Tracking back the satellites to their progenitors, we identify at the end of reionization 23 of the satellite's progenitors to be able to host star formation for both the MW and M31, corresponding to 19 and 15 per cent of the total number of present-day satellites, respectively. Since the timing of reionization and the conditions necessary for star formation (both the minimum halo mass and ionization state, see section~\ref{sec:sources}), this percentage is source model dependent. For these numbers, we include all progenitor haloes larger than $\sim\!10^8\,\msun$ before $z = 6.5$.

\section{Results}

\subsection{Global results}
\label{sec:global}

\begin{figure*}
\begin{center} 
\includegraphics[height=1.6in]{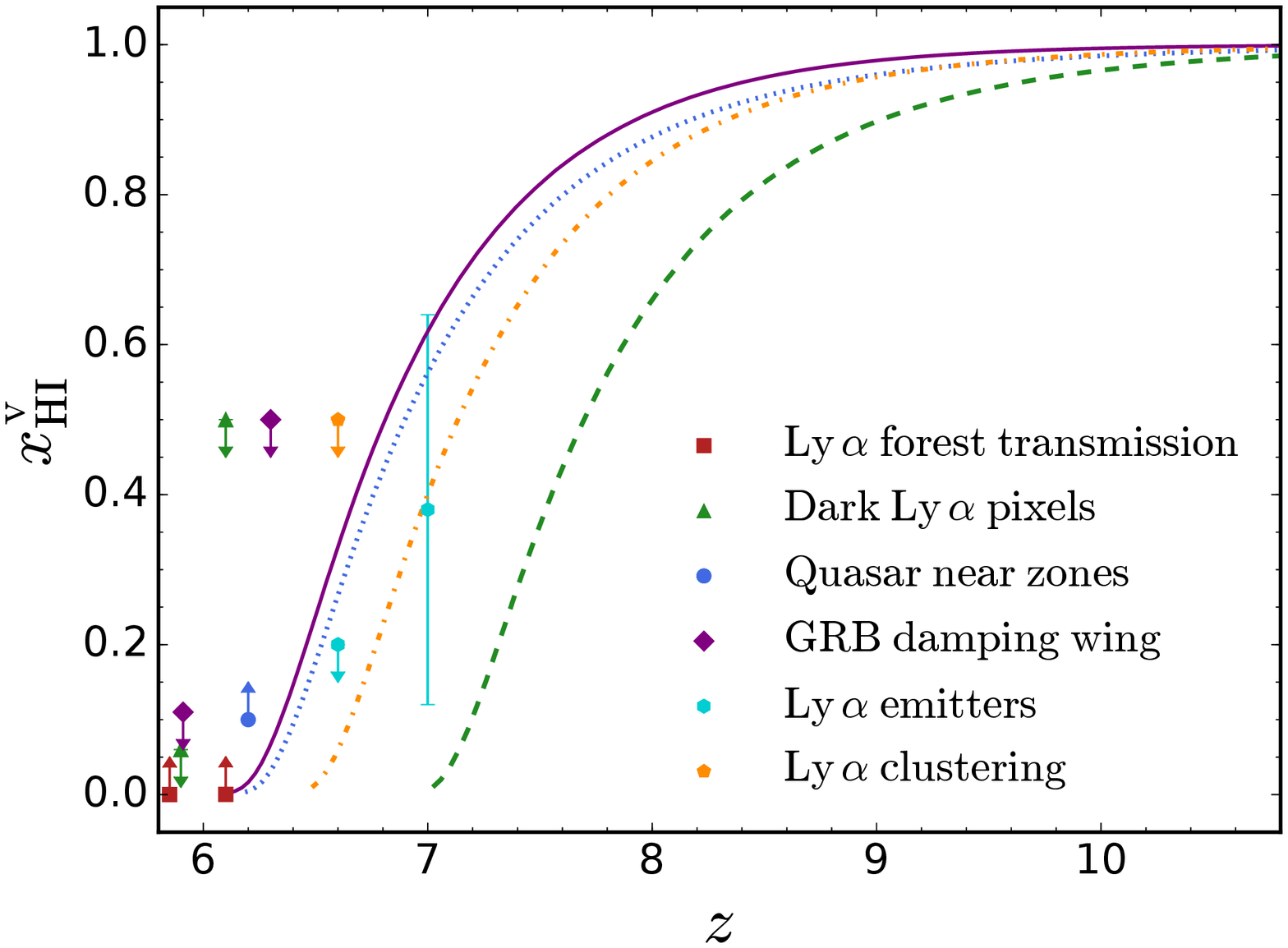} 
\includegraphics[height=1.6in]{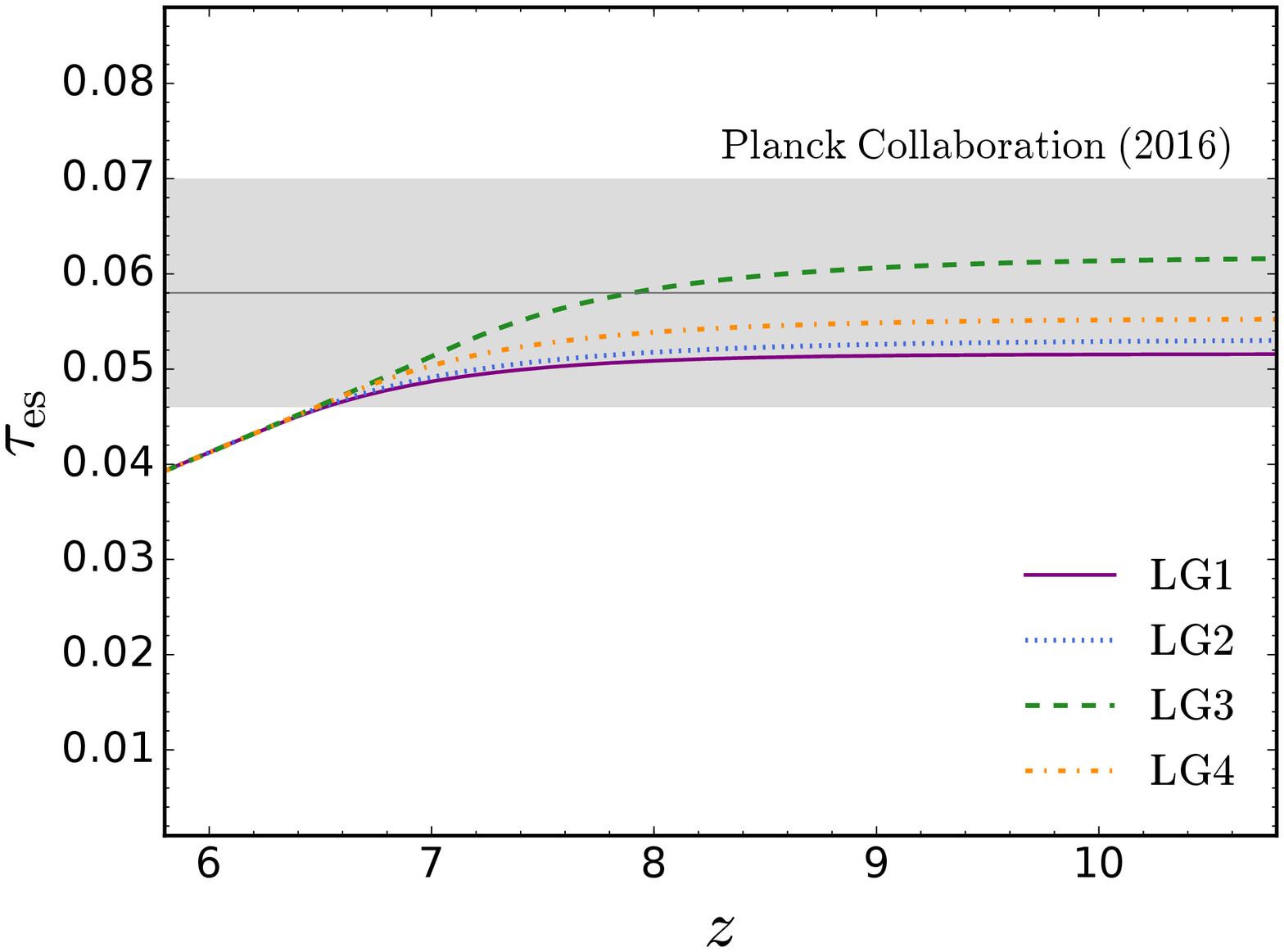}
\includegraphics[height=1.6in]{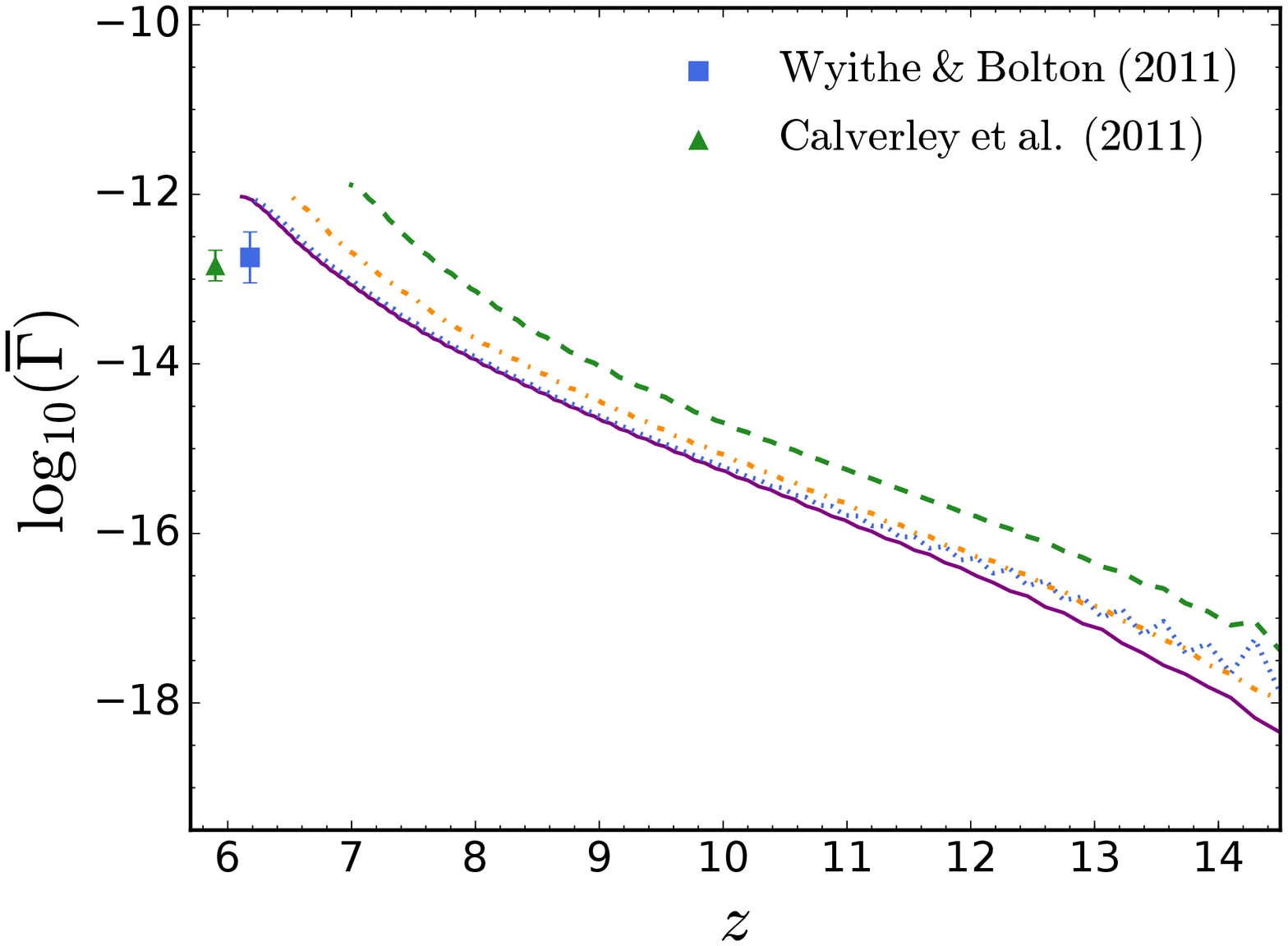}
\caption{The four source models in the 91~Mpc box compared to observational constraints. The solid, dotted, dashed and dot-dashed lines indicate LG1, LG2, LG3, and LG4, respectively. Left: the volume-weighted mean neutral fraction of hydrogen compared to observational inferences from \Lya~forest transmission \citep[squares, red;][]{Fan06}, dark \Lya~ forest pixels \citep[triangles, green;][]{McGr11a,McGr15a}, quasar near zones \citep[circles, blue;][]{Schr13a}, GRB damping wing absorption \citep[diamonds, magenta;][]{McQu08a,Chor13a}, decline in \Lya~ emitters \citep[hexagons, cyan;][]{Ota08a,Ouch10a}, and \Lya~ clustering \citep[pentagons, orange;][]{Ouch10a}, following the discussion in \citep{Robe15a}. Middle: The integrated electron-scattering optical depth compared to the \emph{Planck}TT+lowP+lensing+BAO 2016 results (thin horizontal line) and the 1$\sigma$  error interval (shaded region) \citep{Plan16a}. \emph{Right:} The mean volume-weighted hydrogen photoionization rate compared to the observational constraints \citet{Calv11a} and \citet{Wyit11a} as the (green) triangle and (blue) square, respectively.
\label{fig:obs}}
\end{center}
\end{figure*}

First, we examine the global results from our simulations. The major observables include the integrated electron-scattering optical depth derived from the CMB polarization power spectra, $\tau_{\rm es}$, which suggests an extended process \citep[e.g.][]{Robe15a}, and observations of the IGM and galaxies near the end of reionization, which imply reionization completion by $z\sim6$ \citep[e.g.][]{McGr15a}. Our simulations yield a range of results, generally consistent with these constraints (Fig.~\ref{fig:obs} and Table~\ref{tab:summary}). 

The left-hand panel of Fig~\ref{fig:obs} shows the volume-weighted mean neutral fraction of hydrogen, $x_{\rm \ion{H}{i}}^{\rm v}$, from a variety of observations. The $x_{\rm \ion{H}{i}}^{\rm v}$ derived from measurements of the effective optical depth evolution of the \Lya~forest (including higher-order transitions, if available) are represented by (red) squares \citep{Fan06}. Interpreting this \Lya~transmission as a neutral fraction requires significant modeling, so the resultant neutral fraction is somewhat uncertain \citep{Mesi10}. Upper limits on the neutral fraction from the fraction of dark pixels in the \Lya~forest are more model-independent and are displayed as (green) triangles \citep{McGr11a,McGr15a}. Rare gamma-ray burst (GRB) damping wings provide upper limits shown as (purple) diamonds \citep{McQu08a,Chor13a}. Near-zone sizes around quasars give some information on the minimum neutral fraction (blue, circles), but these measurements are dependent on uncertain intrinsic quasar properties \citep{Bolt11a,Schr13a}. \Lya~emitters \citep{Ota08a,Ouch10a} and their clustering \citep{Ouch10a} provide further constraints shown as (cyan) hexagons and (orange) pentagons, respectively. Our later reionization models (LG1, LG2, and LG4) agree well with the quick rise observed in the neutral hydrogen fraction from $z\sim6-7$. The early reionization model (LG3) is in mild tension with these constraints, given its more numerous sources leading to an earlier end of reionization. Tuning down the source efficiencies in this model would bring the neutral fraction into agreement in a simple manner.

The $\tau_{\rm es}$ calculated from our simulations is listed in Table~\ref{tab:summary} and plotted in the middle panel of Fig.~\ref{fig:obs}. The latest constraints from \emph{Planck}TT+lowP+lensing+BAO data give $\tau_{\rm es} = 0.058\pm0.012$ \citep{Plan16a}, represented by the shaded region in Fig.~\ref{fig:obs}. All models agree within $1\sigma$ or better of the measured values. Finally, the right-hand panel of Fig.~\ref{fig:obs} shows the volume-averaged hydrogen photoionization rate, $\Gamma$. All our simulations predict $\Gamma\sim10^{-12}\,\mathrm{s}^{-1}$ at the end of their respective reionization. Observations (blue square and green triangle) find a lower value of $\Gamma_{\rm obs}=10^{-13}-10^{-12.4}$~s$^{-1}$ at $z \sim 6$ \citep[][respectively]{Calv11a,Wyit11a}. Once again the LG3 model is an outlier, the least in agreement with the data.  
The discrepancy between our results and the observations could be alleviated by including small-scale gas clumping and Lyman-limit systems, which limit the mean free path of ionizing photons, neither of which is included in the simulations presented here. This clumping delays the late stages of reionization, while not having a significant effect on the optical depth \citep{Mao17a}. Similarly, the Lyman-limit systems delay reionization, resulting in slightly lower integrated electron-scattering optical depth and can decrease the mean photoionization rates at the end of reionization by factor $\sim\!3$ \citep{Shuk16a}.

\subsection{Local results}
\label{sec:local}

Beyond the global results, many observations of the Local Universe exist. In this section, we investigate the reionization history of the LG and Virgo. We also look specifically at the MW and M31 and their substructure, where observations of the SFHs of the nearby dwarf galaxies are relevant.

\begin{figure*}
  \begin{center}
    \includegraphics[width=6.78in]{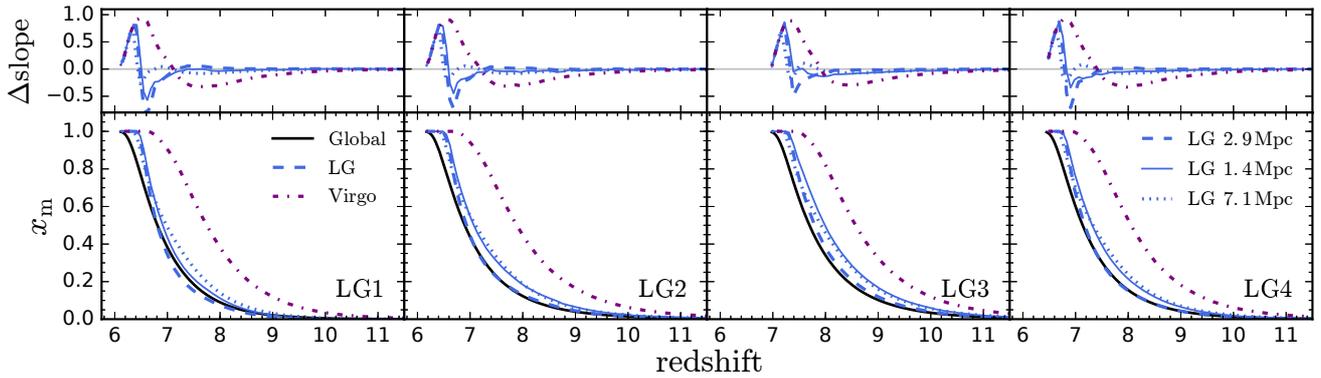}
      \end{center}
  \caption{The mass-weighted ionization fraction for the LG, Virgo, and the entire box for all source models in the lower panels. Simulations LG1, LG2, LG3, and LG4 are shown from left to right, respectively. Plotted are the global result (thick, black line), Virgo, which reionizes earlier (dot-dashed, purple line), and the LG neighbourhood (blue) within 2.86, 1.43, and 7.14~Mpc from the $z=0$ barycentre (dashed, thin, and dotted lines, respectively, as labelled). The upper panels are the differences slope with respect to the global result, which corresponds to the line at zero. Negative values indicate faster reionization than the box as a whole.}
    \label{fig:local_xfrac}
\end{figure*}

We calculate the reionization history using the Lagrangian mass distribution for each object, defined as all the mass that will eventually, by $z=0$, end up within 2.86~Mpc of the barycentre of the LG and the centre of mass of Virgo. This distance was chosen to roughly correspond to the \citet{McCo12a} convention for inclusion in the LG. The resulting mass-weighted ionization fractions, $x_{\rm m}$, as a function of $z$ for LG1, LG2, LG3, and LG4 are shown in Fig.~\ref{fig:local_xfrac}, from left to right in the lower panels. The global result for the entire box is shown (black, solid) for comparison to the $x_{\rm m}$ for LG (blue, dashed) and Virgo (purple, dot-dashed). In all scenarios, the reionization of Virgo occurs earlier than average overall, because (proto) Virgo is within a significantly overdense region, which strongly biases haloes to cluster in the same region. The Virgo reionization curve appears very similar in shape to the global reionization history, thus its reionization is likely internal (i.e., by its own sources), as well.

Conversely, the LG reionization history differs from that of the simulation box and Virgo. When there are no low-mass sources as in LG1, the LG begins reionizing much later as there are few high-mass haloes in this region. For the other three models, the more numerous LMACHs are the dominant producers of ionizing photons, at least initially. The reionization of the LG largely tracks the global one until around 40 per cent ionization, after which the LG reionizes more quickly.  We quantify this behaviour in the upper panels of Fig.~\ref{fig:local_xfrac} by showing the difference in slope between the LG and Virgo as compared to the entire box, where $\Delta$slope$\,\equiv\,d\,\left(x_{\rm m}^{\rm LG, Virgo}\right)/dz - d\,\left(x_{\rm m}^{\rm global}\right)/dz$. The global result is represented as the thin line at zero by definition. Essentially, negative values indicate reionization occurring faster than the box as a whole. In all low-mass-halo cases for the LG, the evolution starts faster than the global one, then slows down and grows slower than the global ionization rate, accelerating again to faster than global at later times. The positive bumps at the end indicate earlier end to reionization. These results provide evidence that the LG is somewhat reionized externally, with a global ionization front swiping quickly through the (proto) LG. 

Statistically, on average the local reionization redshift correlates closely with the density of that region \citep{Ilie06a}. In order to quantify this in the context of the LG, we calculate the local dimensionless density by taking the average density of the cells that reside in the Lagrangian volume that will end up within certain distance from the LG or Virgo divided by the mean global density. Resultant reionization histories are shown in Fig.~\ref{fig:local_xfrac}. For region of 1.43~Mpc around the present-day LG, the dimensionless density of the Lagrangian volume that will become the LG, $\Delta^{\rm LG}_{\rm 1.43~Mpc}$ is 0.9459 at $z = 16.095$, decreasing to 0.9100 by $z = 6.483$. For a 2.86~Mpc region around the LG, $\Delta^{\rm LG}_{\rm 2.86~Mpc}$ starts at 0.9317 at $z = 16.095$ and decreases  to 0.8824 by $z = 6.483$ and for a 7.14~Mpc region around the LG, $\Delta^{\rm LG}_{\rm 7.14~Mpc}$ approaches the average for the Universe, evolving from 0.9831 at $z = 16.095$ to 0.9717 at $z = 6.483$. Thus, the LG lies in a moderately underdense region. For all radii, the LG region ends being reionized very fast compared to the mean, implying a significant outside influence from an ionization front. In contrast, Virgo forms in an overdense region, with overdensity within 2.86~Mpc of $\Delta_{\rm 2.86~Mpc}^{\rm Virgo}$ = 1.0222 (1.0548) for $z = 16.095$ (6.483), which is reflected in its significantly earlier (self-)reionization. 

\subsection{Reionization history of the MW, M31, and their satellites}
\label{sec:reion}

In this section, we investigate the intrinsic and environmental effects that drive reionization history of individual objects, including the MW, M31, and their satellites as identified in section~\ref{sec:sats}. Most directly, the more massive an object is, the higher its star formation potential. Beyond halo mass, the reionization history of an object can affect its SFH. As outlined above in section~\ref{sec:intro} and \ref{sec:sources}, dwarf galaxies -- in this case, LG progenitors -- are susceptible to radiative feedback, whereby an ionized region may have suppressed star formation. 

\begin{figure}
  \begin{center}
    \includegraphics[width=3.2in]{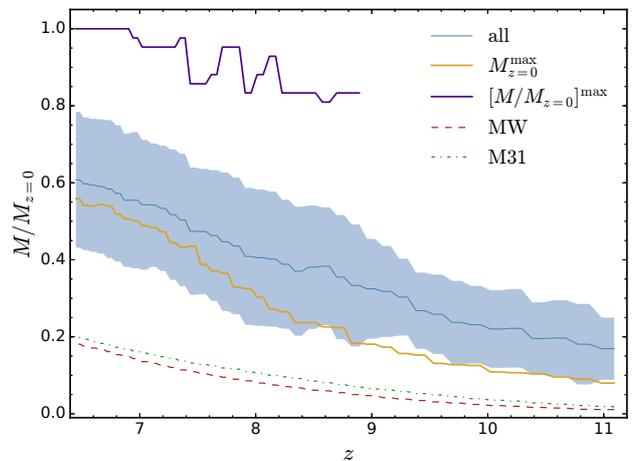}
   \end{center}
  \caption{The evolution of the fraction of mass accumulated into the bound progenitors of the MW and M31 satellites during reionzation. The thin line represents the average value, and the shaded region indicates the 1$\sigma$ spread. The $M^{\rm max}_{z=0}$ (middle, yellow) and [$M/M_{z=0}]^{\rm max}$ (upper, purple) satellites are the thick lines. The lowest lines are the mass fractions of the MW (dashed) and M31 (dot-dashed).}
    \label{fig:mvz}
\end{figure}

\begin{figure*}
  \begin{center}
   \includegraphics{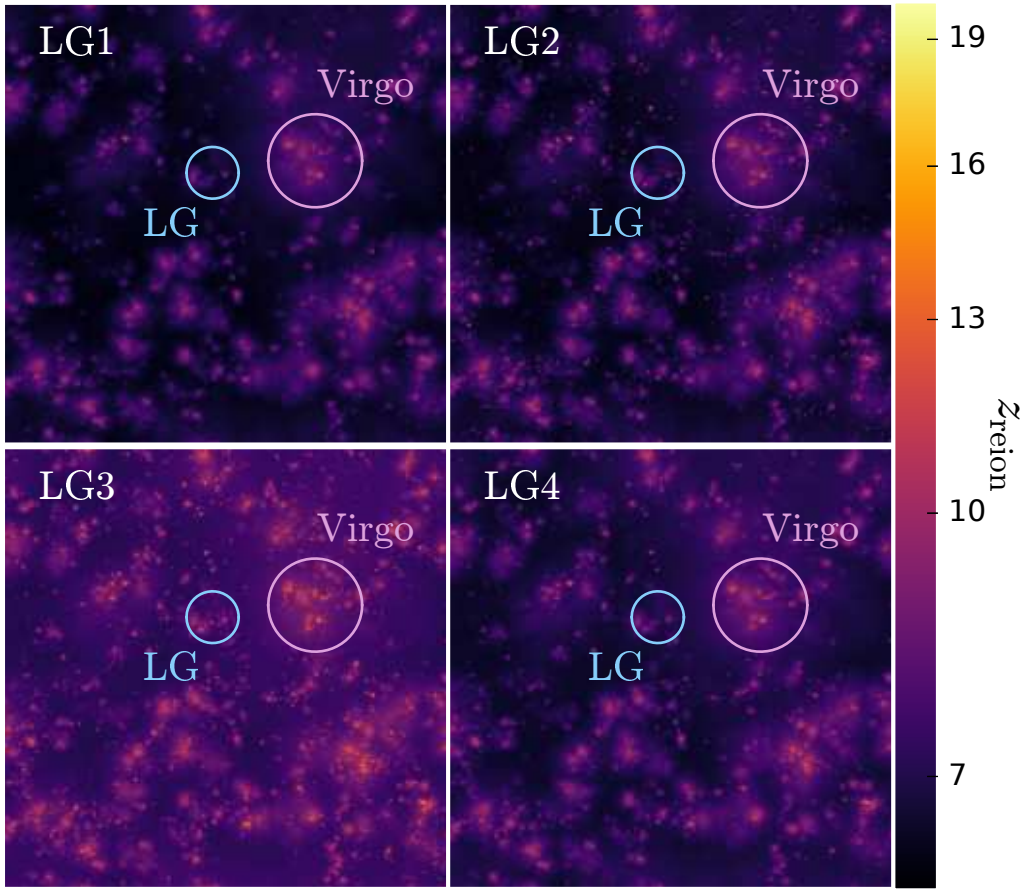}
  \end{center}
  \caption{Map of the reionization redshifts throughout our volume for models LG1 (upper left), LG2, (upper right), LG3 (lower left), and LG4 (lower right). The slice depth is 0.36~Mpc (a single pixel width), and the entire box is shown, making each side 91~Mpc. The small (blue) and large (purple) circles approximately represent the Lagrangian volume that will end up within 2.86~Mpc of the LG and Virgo, respectively.}
    \label{fig:image_zr}
\end{figure*}

Directly relevant to this work is how much of the present-day mass of a DM halo hosting a central galaxy or its satellites resides within bound structures before the end of reionization, as these progenitors may be able to form stars (depending on the source model and the local environment at the time). In Fig.~\ref{fig:mvz}, we show the total mass in DM haloes above $10^8~\msun$ that eventually ends up in the MW (red, dashed), M31 (green, dot-dashed), or their satellites in units of present-day mass, $M_{z = 0}$. By the end of the EoR, the MW and M31 have fractions of accumulated mass, $M/M_{z=0}$, that are nearly 20 per cent. Therefore, significant star formation occurs in the central galaxies before the end of reionization, as expected. 

In Fig.~\ref{fig:mvz}, we also plot the average $M/M_{z=0}$ and 1$\sigma$ scatter for the satellites, as the thin (blue) line and the shaded region, respectively. At $z = 0$, we find 120 satellites within the MW DM halo with a mass range of $10^{7.94} - 10^{9.69}\,\msun$. At $z \sim 6.5$, 23 MW satellites have progenitors that have accumulated enough DM to possibly form stars. Similarly for M31, we find 151 satellites at $z=0$ with a mass range of $10^{8.02} - 10^{9.54}\,\msun$. As a slightly larger DM halo, more substructure is expected; however, the largest satellite is smaller than the MW equivalent. At $z \sim 6.5$, 23 M31 satellites have progenitors that have sufficient mass to possibly form stars. Generally, the satellites increase in mass fraction over time \footnote{Note that we are only considering \emph{surviving} satellites. Satellite progenitors can also lose mass over time due to environmental effects.}, and the mass fraction at the end of reionization exhibits significant scatter. By the end of reionization, the satellites present have accumulated, on average, $\sim\!60$ per cent of their final mass. 

The most unusual individual cases are plotted separately. The lower (thick, yellow) line represents the largest present-day satellite, which happens to reside in the MW halo and possibly an object akin to a Magellanic cloud. Initially, its mass fraction lags behind the average, increasing only gradually, similarly to the much larger MW. Around $z \sim 8$, the mass accumulation accelerates, and this largest satellite ends up with a nearly average $\left[M/M_{z=0}\right] \approx 0.55$. The upper (thick, purple) line shows one of the smallest satellites that has all of its mass already bound into a progenitor by the end of reionization, or $\left[M/M_{z=0}\right] = 1.0$. This satellite also belongs to the MW, and loses mass between this time and $z = 0$.

The ionization state of a halo plays a major role, along with its mass, in determining the amount of star formation occurring within it. We define $z_{\rm reion}$ as the redshift at which the bound mass exceeds a mass-weighted ionization threshold of 0.5\footnote{The exact threshold chosen slightly impacts the overall behaviour, as a lower value will generally increase $z_{\rm reion}$ and individual satellites may shift. No trends or correlation change significantly, especially since the satellites tend to ionize quickly. We choose the first slice at which the threshold is met.}. In Fig.~\ref{fig:image_zr}, we show $z_{\rm reion}$ for a 0.63~Mpc thick slice (equivalent to a single cell width) through our simulation box, where the lighter regions reionize earlier. The approximate Lagrangian volume (at $z = 6$) for the LG and Virgo progenitors are marked by the small (blue) and large (purple) circles, respectively. The LG1, LG2, LG4, and LG3 are displayed clockwise, starting with the top-left panel. Immediately obvious is the fact that LG3, with its higher efficiencies and large number of sources, reionizes the earliest. Also, in all models, the Virgo progenitors tend to be some of the earliest reionized regions. The LG as a whole does not reionize especially early, though this result is not spatially uniform. 

\begin{figure*}
  \begin{center}
    \includegraphics[width=6.7in]{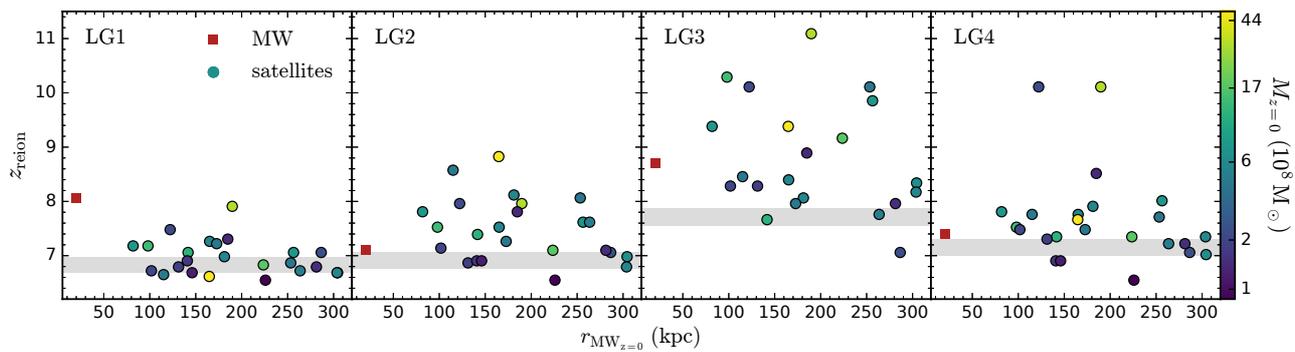}
  \end{center}
  \caption{The reionization redshift as a function of the present-day distance to the MW. The colour of the points represent the present-day halo mass for MW satellites from darkest (purple) to lightest (yellow) for the least to most massive. Simulations LG1, LG2, LG3, and LG4 are shown from left to right, respectively. The (red) square near $r_{\textrm{\scriptsize MW}}=0$ is $z_{\rm reion}$ for the MW. The shaded region indicates $x_{\rm m} = 0.4 - 0.6$.}
    \label{fig:sats_zrm}
\end{figure*}

\begin{figure*}
  \begin{center}
    \includegraphics[width=6.7in]{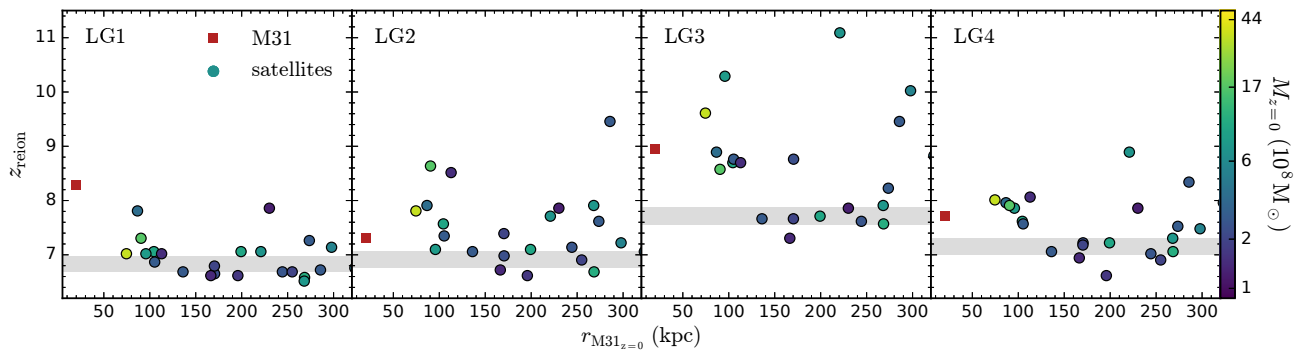}
   \end{center}
  \caption{Same as Fig.~\ref{fig:sats_zrm}, but for M31.}
    \label{fig:sats_zrm_M31}
\end{figure*}

The MW and M31 satellites' progenitors may have a reionization history distinct from the average of the LG volume. In Fig.~\ref{fig:sats_zrm}, we plot $z_{\rm reion}$ of the satellites, compared to their present-day distance from the central galaxy, $r_{\textrm{\scriptsize MW}_{z=0}}$. The darkest (purple) to lightest (yellow) colour dots are the least to most massive $M_{z=0}$, respectively. The grey band represents the global $x_{\rm m} = 0.4 - 0.6$, where the LG reionization accelerates indicating external influence (c.f. Fig.~\ref{fig:local_xfrac}). The four source models LG1, LG2, LG3, and LG4 are shown from left to right, as labelled. In the models with their more numerous small sources, there is some indication that satellites nearer to the MW reionize earlier. For all models, the external sweep of reionization accounts for the reionization of a significant fraction of the satellites, meaning many satellites will have similar $z_{\rm reion}$ within the shaded band. Note that several satellite progenitors at roughly the same present-day distance can have vastly different $z_{\rm reion}$.

The $z_{\rm reion}$ for the MW's own progenitors is shown as the (red) square. For LG1, the MW reionizes before any of the satellite progenitors. For LG3, the MW reionizes at an average time compared to the satellites. For LG2 and LG4, the MW reionizes right around the time that reionization of the LG accelerates (the shaded band), along with most of its satellites. Though not shown, we find that the closer a satellite is to the MW at the satellite's $z_{\rm reion}$ has no bearing on its reionization redshift. Essentially, the MW progenitors are not the main sources of ionizing photons for its satellites, as most satellite progenitors reside farther than 1~Mpc away.

We also find that $z_{\rm reion}$ is only loosely correlated with the satellite mass, with the more massive (lighter circles) satellites generally reionizing earlier. Some large satellites have progenitors capable of reionizing themselves or were near other large structures. This trend may not be true for specific satellites. For a given $M_{z=0}$, $z_{\rm reion}$ shows significant scatter. These general conclusions are model-independent, though the exact $z_{\rm reion}$ for each satellite is different for the four source models. Almost by definition, the earlier a model globally reionizes, the higher $z_{\rm reion}$ for the satellites. In Fig.~\ref{fig:sats_zrm_M31}, we show the analogous results to Fig.~\ref{fig:sats_zrm}, but now for M31. The results are very similar to the MW case. Overall, there is less scatter in $z_{\rm reion}$ for the M31 satellites. Also, M31 reionizes earlier than the MW, which may be due to the fact that our M31 is closer to Virgo during reionization and/or our M31 is more massive.

\citet{Ocvi14a} found a correlation between $z_{\rm reion}$ for a satellite and the present-day distance from the central galaxy in some of their source models. Their methods are somewhat similar, using a constrained $N$-body simulation and an RT post-processing (more simplified than our methods), but in a smaller volume at higher resolution. We identify fewer satellites, which is expected as our minimum present-day mass threshold for inclusion as a satellite is higher, and this smaller set may mean we miss trends present in their data. We find only a slight correlation between $z_{\rm reion}$ and the present-day distance from the central galaxy, weaker than that of \citet{Ocvi14a}. Our results are all consistent with zero slope, especially in the M31 case. Therefore, we do not confidently predict a relation between $z_{\rm reion}$ and present-day distance from the central galaxy for satellites; though, we address a similar correlation in the next section.

\subsection{Ancient star formation in the satellites of the MW and M31}
\label{sec:SF}

Within our models, a higher halo mass does not necessarily imply more star formation. Depending on the ionization state of the cell containing an LMACH source and the source model, star formation may be partially or fully suppressed. Furthermore, although the measured quantity is the present-day mass of satellites, the historical accumulation of mass determines the potential star formation prior to the end of reionization (see section~\ref{sec:reion}). Because of these considerations, the environmental and mass accretion history of the satellites influence SFH, and these quantities are only loosely related to present-day satellite properties, such as distance from the MW or M31 and the DM halo mass. Although we do not consider them here, other baryonic effects can also impact the SFH and the stellar mass fraction.

\begin{figure*}
  \begin{center}
    \includegraphics[width=6.7in]{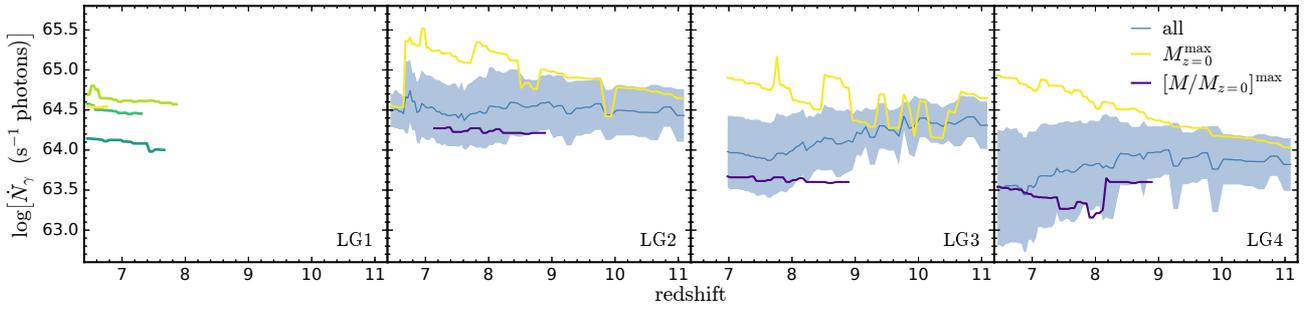}
  \end{center}
  \caption{The ionizing photon emissivity of the progenitors for all MW satellites during reionization. For satellites with multiple progenitors, $\dot{N}_{\gamma}$ is summed. Simulations LG1, LG2, LG3, and LG4 are shown from left to right, as labeled. For the leftmost panel, the most to least massive satellites (at $z = 0$) are lightest to darkest. For the rest, the thin line represents the average value, and the shaded region indicates the 1$\sigma$ spread. For reference, the $M^{\rm max}_{z=0}$ (yellow) and [$M/M_{z=0}]^{\rm max}$ (purple) satellite are the thick lines. The results of our suppression models is seen in the decrease at later times, or higher ionized fractions. }
    \label{fig:sats_raw}
\end{figure*}

\begin{figure*}
  \begin{center}
    \includegraphics[width=6.7in]{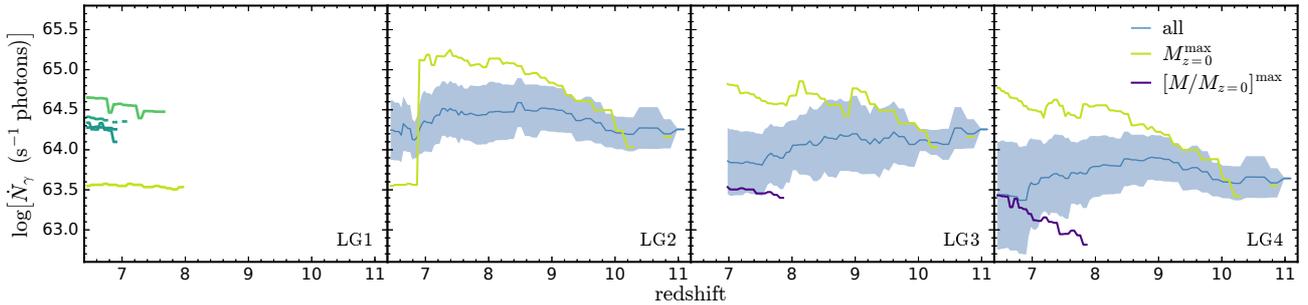}
  \end{center}
  \caption{ Same as Fig.~\ref{fig:sats_raw} for M31 satellites.
}
\label{fig:M31_raw}
\end{figure*}

Strictly speaking, the quantity `measured' from our simulations is the ionizing emissivity of photons, or $\dot{N}_{\gamma}$ as defined in equation~(\ref{eq:Ndot}). In Fig.~\ref{fig:sats_raw}, $\dot{N}_{\gamma}$ as a function of $z$ is plotted for models LG1, LG2, LG3, and LG4 from left to right. For LG1, the individual satellites are shown. For the rest of the models, the average value is the thin line and the shading represents the 1$\sigma$ scatter. Here, $g_{\gamma}$, which is a component of equation~(\ref{eq:Ndot}), is determined by the ionization state of the cell in which the progenitor resides and the mass of that halo. This halo mass need not be the same as the number of progenitor particles that end up in the satellite, due to later stripping and mergers. Therefore, a progenitor -- and a given satellite may have many progenitors -- may have a higher (or lower) efficiency than dictated by the total accumulated mass. $\dot{N}_{\gamma}$ exhibits significant scatter that increases at later times with more satellite progenitors becoming massive enough to become star-forming and with the increasing influence of source suppression.

For LG1 in the leftmost panel, four satellite progenitors are large enough to form stars in this model. These progenitors all reside in haloes $>\!10^9\,\msun$, which are impervious to radiative feedback in our model, and can thus continue forming stars after reionization. The lighter (yellow) lines represent more massive haloes than than the darker (teal) lines. Notably, the most massive present-day satellite only has the third highest $\dot{N}_{\gamma}$ at the end of reionization. For LG2 in the centre-left panel, the average value is fairly flat, and the end of reionization is clear from the sharp drop at $z \sim 6.7$. At this time, most satellite progenitors are suppressed and incapable of producing ionizing photons, reducing the scatter. For LG3 in the centre-right panel, the effect of suppression is seen at earlier times with progenitors exhibiting high efficiency and, at later times, decreasing in efficiency even though the accumulated mass is increasing. For LG4 in the rightmost panel, late times show more than an order-of-magnitude scatter, and the average value decreases over time due to source suppression.

Similar to Fig.~\ref{fig:mvz}, we show two special cases with $M_{z=0}^{\rm max}$ (upper, yellow line) and $\left[M/M_{z=0}\right]^{\rm max}$ (lower, purple line). Especially in the $M_{z=0}^{\rm max}$ progenitors, the effect of suppression is clear, where the region of one or more of the progenitors becomes ionized and $\dot{N}_{\gamma}$ decreases. As the environment changes, more mass accretes, or more progenitors form, $\dot{N}_{\gamma}$ may subsequently increase again. This trend is most apparent in LG2 and LG3 (middle two panels), as those models feature sharp suppression of star formation. For $\left[M/M_{z=0}\right]^{\rm max}$, $\dot{N}_{\gamma}$ is quite flat, but the exact history is model-dependent. For example, this satellite's progenitor has evidence of suppression in LG4 (rightmost panel) and becomes fully suppressed by the end of reionization in LG2.

In Fig.~\ref{fig:M31_raw}, the same quantity is plotted for the satellites of M31. For LG1 in the leftmost panel, the largest satellite today produces the fewest ionizing photons during reionization. This particular satellite has many smaller progenitors, but just one large progenitor that does not significantly increase in mass during reionization. Similarly to the MW, the LG2 model (in the centre-left panel) shows evidence of an ionization front, which occurs slightly earlier for M31 than MW. In the centre-right panel, LG3 exhibits significant scatter throughout reionization, and a gradual decrease at later times. Once again similarly to the MW, LG4 (rightmost panel) shows larger than order-of-magnitude scatter, especially at late times.

The special cases of $M_{z=0}^{\rm max}$ (upper, yellow line) and $\left[M/M_{z=0}\right]^{\rm max}$ (lower, purple line) are again included. For the largest satellite, the effect of suppression is especially pronounced in LG2 (centre-left panel), causing a large drop in $\dot{N}_{\gamma}$ that coincides with the dip for all satellite progenitors. More interesting for M31 is the $\left[M/M_{z=0}\right]^{\rm max}$ satellite. For both LG1 and LG2 (left two panels), no star formation occurs in this satellite before the end of reionization, i.e. not present in the plot. For the other models, $\dot{N}_{\gamma}$ for this satellite is mostly flat in LG3 and steadily increases in LG4. These differences demonstrate that an individual satellite's history can vary significantly depending on the source model. 
 
These `simulation units' for the star formation activity are not directly observable; therefore, we make a few additional assumptions to convert to observable quantities. Using equation~(\ref{eq:eff}), the total mass of stars formed during a time-step is
\be
\textrm{SF} ~~=~~ f_*M_{\rm prog}\frac{\Omega_0}{\Omega_{\rm B}} ~~=~~ \frac{g_\gamma M_{\rm prog}}{f_{\textrm{esc}} N_{\rm i}}\frac{10\,\rm{Myr}}{t_{\rm s}}\frac{\Omega_0}{\Omega_{\rm B}},
\label{eq:SF}
\ee
where $M_{\rm prog}$ is the progenitor mass and $g_\gamma$ and $N_{\rm i}$ are dependent on the source model and the mass of the halo in which the progenitors reside (see section~\ref{sec:sources}). Given that the escape fraction is a highly uncertain quantity, varying widely in both observations and simulations, we simply adopt $f_{\rm esc} = 0.1$ as a reasonable intermediate value for all haloes (c.f. \citet{Ma15a, Xu16a} for recent numerical studies). For low-efficiency haloes, we choose a constant $N_{\rm i}^{\rm low} = 4\times10^3$, which is consistent with a Salpeter IMF \citep{Leit99a}. For high-efficiency haloes, we assign a higher $N_{\rm i}^{\rm high} = 1\times10^4$, consistent with a somewhat more top-heavy IMF. Essentially, high-efficiency, low-mass haloes are assumed to be more efficient producers of ionizing photons.

Depending on the source model, the number of satellites with actual star formation before the end of reionization is 23 (19 per cent) in LG4, 20 (17 per cent) in LG3 and LG2, and just 4 (3 per cent) in LG1. The fact that LG2 and LG3 have the same number is simply a coincidence, as some satellites are immediately and completely suppressed in LG2 and some satellite progenitors do not form in LG3 before the end of its earlier reionization. We do not consider any star formation or possible disruption of satellites post-reionization, as we identify haloes present at $z=0$ and track them back in time. In a model with aggressive suppression (LG2), only four satellites can maintain star formation through the end of reionization. This number is not coincidentally the same as the number star-forming satellites in LG1; there are the only four satellites with progenitors that are greater than $10^9\,\msun$ before the end of reionization. 

\begin{figure*}
  \begin{center}
    \includegraphics[width=6.7in]{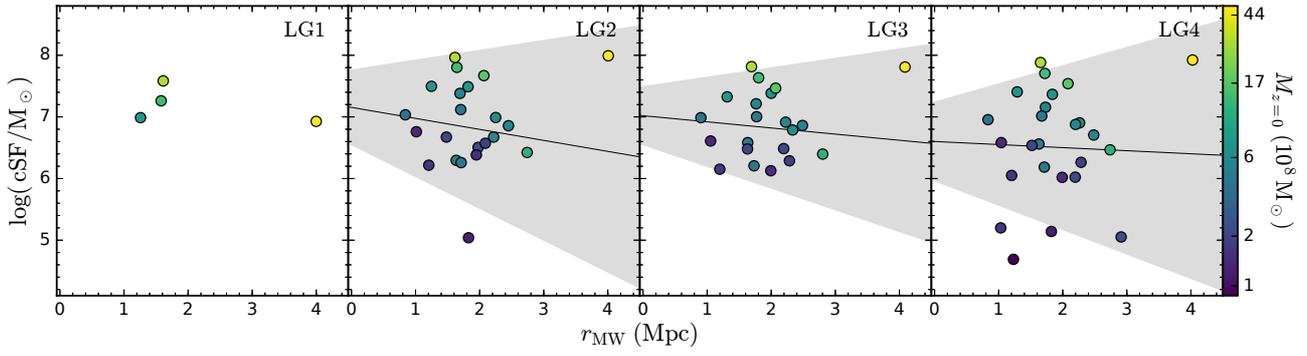}
    \vspace{-3mm}
  \end{center}
  \caption{The cumulative star formation in MW satellites in units of $\msun$ as a function of the progenitors distance from the centre of mass of the MW progenitors. Simulations LG1, LG2, LG3, and LG4 are shown from left to right. The colour of the circles represent the present-day halo mass for MW satellites from darkest (purple) to lightest (yellow) for the least to most massive (see colour bar). The thin line represents the ordinary least squares regression with the shaded region indicating the 68 per cent confidence interval. Only in this plot, the most massive satellite is treated as an outlier and removed from this analysis.}
    \label{fig:mstar_r}
\end{figure*}

In Fig.~\ref{fig:mstar_r}, we show the cumulative star formation (cSF) in units of $\msun$ for each MW satellite in each of our source models as a function of distance, $r_{\textrm{\scriptsize MW}}$, from the MW at $z_{\rm reion}$ for each satellite. The lightest (yellow) to darkest (purple) circles indicate the most to least massive, respectively, present-day satellites. Generally, the more massive (lighter in colour) satellites have higher cSF.  This correspondence is not exact however, especially for the smallest satellites that are most susceptible to suppression or a break up of their dark matter halo. In particular, the fifth largest satellite has some of the lowest star formation in models LG2 and LG3, where suppression can have a major impact. This same satellite lies more in the middle of the pack for LG4 and is not present at all for LG1.

For the three models with LMACHs, we have also included the ordinary least squares fit to data, shown as a thin line in Fig.~\ref{fig:mstar_r}. The shaded region represents the 68 per cent confidence interval. Only for this plot, the most massive (which is also farthest afield during reionization) satellite is treated as an outlier and removed from the analysis. Given the small number of points, the result is mainly a scatter plot, unsurprisingly. Essentially, for all three models, the slope of the fit is consistent with zero, indicating no relationship between cSF and the distance from the MW progenitors. Since LG1 only has four points, we have foregone this particular analysis.

\begin{figure*}
  \begin{center}
    \includegraphics[width=6.7in]{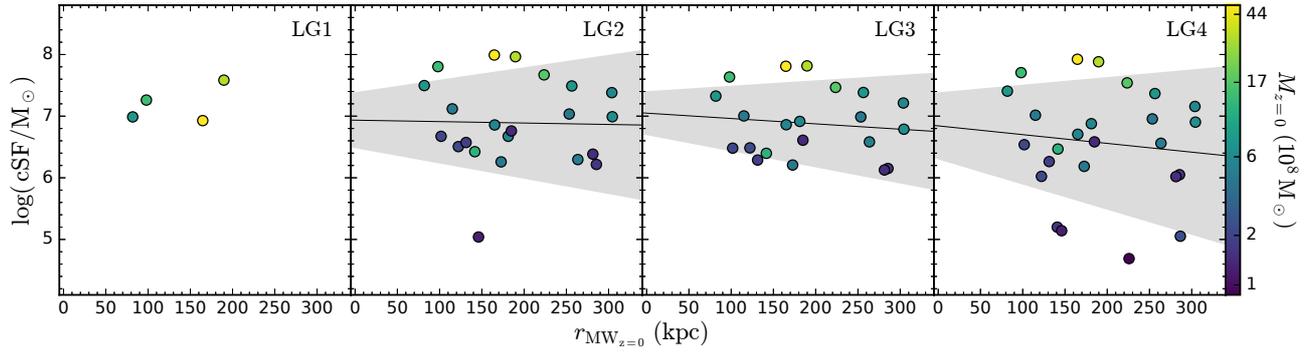}
  \end{center}
   \vspace{-3mm}
  \caption{Same as Fig.~\ref{fig:mstar_r}, but compared to the present-day distance from the MW.
  }
    \label{fig:mstar_r0}
\end{figure*}

In Fig.~\ref{fig:mstar_r0}, we show the cSF of the satellites vs. the \emph{present-day} distance from the MW with all the same features as the previous plot. Note that no satellite was excluded in this fitting analysis. Although the spread in the slope has shrunk, the data are all consistent with no correlation. LG4 may be an exception, showing slightly negative trend within 68 per cent confidence.

\begin{figure*}
  \begin{center}
    \includegraphics[width=6.7in]{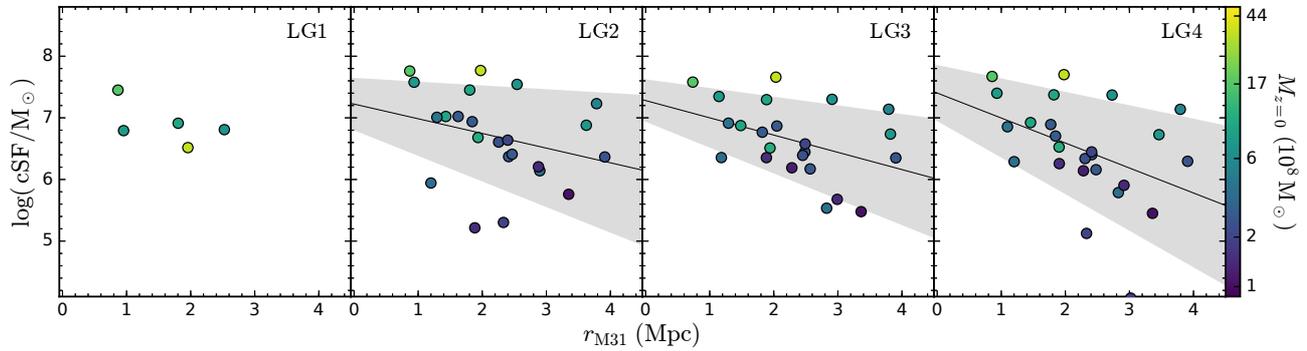}
  \end{center}
   \vspace{-3mm}
  \caption{Same as Fig.~\ref{fig:mstar_r} for M31 satellites.
  }
    \label{fig:mstar_r_M31}
\end{figure*}

\begin{figure*}
  \begin{center}
    \includegraphics[width=6.7in]{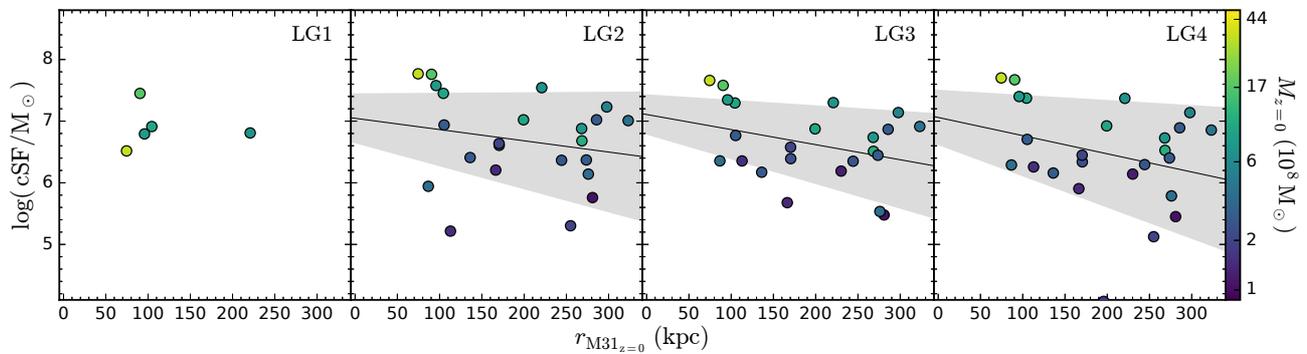}
  \end{center}
   \vspace{-3mm}
  \caption{Same as Fig.~\ref{fig:mstar_r0} for M31 satellites.
  }
    \label{fig:mstar_r0_M31}
\end{figure*}

In Fig.~\ref{fig:mstar_r_M31}, we show cSF for each M31 satellite in each of our source models as a function of distance, $r_{\textrm{\scriptsize M31}}$, from M31 at $z_{\rm reion}$ for each satellite. Some of the same features hold for M31 as the MW. The most massive satellites have the highest cSF, and the least massive satellites -- the ones most susceptible to feedback -- exhibit the most scatter. Interestingly, M31 has the opposite trend from the MW: more massive satellites have smaller $r_{\textrm{\scriptsize M31}}$, so closer satellites have higher cSF. Similar to the MW, the number of satellites with actual star formation before the end of reionization ranges from 23 (16 per cent) in LG4 to 22 (15 per cent) in LG3 and LG2 to 5 (3 per cent) in LG1. In a model with sharp suppression (LG2), only five satellites can maintain star formation through the end of reionization.

In Fig.~\ref{fig:mstar_r0_M31}, the cSF for all M31 satellites are displayed as a function of $r_{\textrm{\scriptsize M31}_{z=0}}$ with the colours corresponding to the present-day mass of the satellites as previously. As before, we have included a linear fit with the 68 per cent confidence level as the shaded region. The trend of decreasing cSF with increasing distance from the central galaxy is stronger for M31 than for the MW. The slope is fairly shallow with a large spread. As before, the correlation tightens when moving from $r_{\textrm{\scriptsize M31}}$ to $r_{\textrm{\scriptsize M31}_{z=0}}$.

Comparing the results for the MW and M31 (Figs.~\ref{fig:mstar_r0} and \ref{fig:mstar_r0_M31}), we see that, although an individual satellite may have a very different cSF in a given model, several general results consistently hold. More massive satellites tend to have more star formation, but the present-day mass of a satellite is not an exact predictor of the stellar mass. The present-day distance of a satellite from their parent has some bearing on the amount of star formation prior to the end of reionization, driven in part by the fact that more massive satellites in M31 tend to be closer to the central galaxy. As this trend is decreasing, the result is not the na\"{\i}ve conclusion drawn from radiative feedback concerns, as in nearer satellites should have their SF more suppressed. When instead looking at cSF/$M_{z=0}$, the correlation is even weaker, though trending positive, and the satellite progenitors tend to reside fairly far from the MW and M31 progenitors. In total, these considerations paint a complicated theoretical picture: the mass, present-day distance from central galaxy, and source model all affect the cSF prior to the end of reionization. Although the MW and M31 live in the same neighbourhood and have similar masses, their reionization signatures differ notably.

\section{Comparison to observations}
\label{sec:obs}

Extensive theoretical and observational discussion of `fossils' resulting from reionization exists in the literature \citep[e.g.][]{Rico05a,Mada08a,Bovi11b}. Although definitions vary, we adopt the convention of \citet{Rico05a} that a fossil is classified as having formed $\gtrsim\!70$ per cent of its stars prior to the end of reionization. Though not an exact correspondence, we will use $M/M_{z=0}$ at $z=6.5$ as a proxy for per cent of stars formed by the end of reionization. Recently, \citet{Bovi11a} estimated that 50 - 70 percent of satellites should be fossils. We find approximately six per cent of satellites for both the MW and M31 -- seven and nine, respectively -- would be classified as fossils, though this fraction is dependent on the source model and on our halo classification thresholds. Of course, LG1 with only large sources and LG2 with aggressive suppression of low-mass haloes will have even smaller fractions. Observationally, \citet{Weis14b} found five fossils out of a sample of 38, equivalent to 13 per cent. Admittedly, the sample is small and by no means comprehensive. Furthermore, recent and ongoing observations that find more ultra-faint dwarf galaxies, \citep[e.g.][]{Bech15a,Kopo15a,Kim15a,Laev15b}, may increase \emph{or} decrease this fraction. For example, \citet{Brow14a} find the SFHs of six ultra-faint dwarf galaxies to be consistent with 80 per cent of their stars forming by $z\sim6$. Generally, we find that our simulated fossils have no distance-to-central-galaxy or mass-dependence, which is consistent with observations. 

Although an exact comparison of our cSF to measured SFHs of LG dwarf galaxies is outside the scope of this paper (we do not track star formation down to $z=0$), we can make some general comparisons to recent observations in the literature. First, \citet{Weis14b} found five LG dwarf galaxies exhibit significant pre-reionization star formation (defined as $>\!90$ per cent stellar mass prior to $z=6$), though early time resolution of derived SFHs is limited and potentially biased \citep{Apar16a}. For a somewhat equivalent measure from our simulation, we find two satellites to have $M/M_{z = 0} > 0.9$ prior to the end of reionization. Since we do not include field dwarf galaxies (left to an upcoming paper) and the exact number is highly mass threshold dependent, we are consistent with these observations. Second, the entire sample of 38 showed evidence of some star formation prior to $z=6$, which is also compatible with our 46 satellites. Of course, the \citet{Weis14b} sample is not comprehensive, so we would not expect exact agreement. 

As for the lack of a widespread reionization signature in \citet{Greb04a} among others, we would argue that the exact `quenching' effect of reionization is uncertain, whether the quenching is permanent, partial, or even universal. These SFHs, however, can rule out certain models. For example, our most extreme suppression model -- LG1 with no star formation below $10^9\,\msun$ -- is ruled out by most observations, because only nine LG satellites would have star formation prior to the end of reionization. For similar reasons, the aggressive suppression model -- LG2 with no star formation below $10^9\,\msun$ in \emph{ionized} regions -- is also unlikely, as only those nine satellites could continue forming stars without significant subsequent accretion. Therefore, just based the existence of ancient stellar populations in dwarf galaxies, one can eliminate certain radiative feedback models. Note that the exact mass threshold can change these conclusions.

Along these lines, \citet{Skil16a} find evidence of a slowdown in star formation after reionization in several M31 satellites. This behaviour is consistent with our LG3 and LG4 models, which both feature partial suppression. The authors also find a satellite exhibiting increased star formation post-reionization, which is also consistent with LG3 and LG4. Both models feature partial suppression, but also feature satellites that are capable of significant star formation post-reionization. On a different note, \citet{Mone16a} carefully study the SFH of Andromeda XVI and find an extremely low-mass galaxy with later quenching that is not coincident with reionization. Though no likely candidates in M31, the MW satellites have at least two satellites with $M_{z=0} \lesssim 10^8\,\msun$ that have accumulated $\lesssim 50$ per cent of their final mass.

Intriguingly, \citet{Skil16a} and \citet{Weis14c} find differences between the SFHs of the MW and M31, albeit in a small sample ($>\!10$ satellites per central galaxy). The main difference is that the M31 sample contains no late-time quenching galaxies, whereas the MW sample has several at comparable luminosities. While our results are not directly applicable to these observed results, we do find some differences between the MW and M31, though our identification is somewhat arbitrary. The fraction of satellites with star formation prior to the end of reionization is higher for the MW. Alternatively, we do find that the M31 satellites are more similar to each other, with more M31 satellites sharing similar $z_{\rm reion}$ and a tighter cSF-$r_{\textrm{\scriptsize M31}_{z=0}}$ relation. \citet{Skil16a} do find more variety of quenching times for MW as compared to M31. Although clearly not conclusive, our results are consistent with observations and point to potentially significant differences in the satellites of the two similarly sized pair galaxies.

\section{Summary}
\label{sec:summary}

We present a suite of simulations designed to investigate the reionization of the Local Group and the potential observable signatures today. To achieve these goals, we use a large (91~Mpc) $N$-body simulation to follow the dark matter evolution and applied full 3D radiative transfer to track reionization. To mimic our Local Universe as closely as possible, we used initial conditions constrained by the observed galaxy peculiar velocities that, by construction, reproduce realistic Milky Way, M31, Virgo (the largest nearby cluster), local void, and Virgo filament. As the exact nature of how reionization proceeds in the early Universe is still uncertain, we employ four source models, where the main distinction is the effect of radiative feedback on ongoing reionization. Essentially, we vary the degree of suppression of star formation in low-mass haloes resulting from reionization. 

We find that the LG mostly reionizes itself with a small amount of influence from its environment, though the overall timing of reionization is dependent on the local environment and source model. We also find that Virgo reionizes itself much earlier than the entire box, as its progenitors reside in a overdense region. M31, nearer to Virgo and more massive, reionizes earlier than the MW. That the reionization of the LG proceeds mostly in an inside-out fashion with some outside influence is further supported by the evolution of the slope of the reionization history, which indicates the rapidity with which the reionization occurs. Global reionization is by definition done by internal sources and serves as the baseline. The Virgo slope is similar to the global one, but shifted to earlier time and reionizing faster, as could be expected for internal reionization of an overdense region. In contrast, the slope for the LG is sharply negative at late times, consistent with partial external reionization, which occurs quickly with the passage of a large-scale ionization front.

The redshift at which the satellite progenitors are reionized, defined as 50 per cent of the bound dark matter residing in ionized cells, is very loosely related to intrinsic or environmental properties. On the whole, more massive satellites reionize earlier. Similar to \citet{Ocvi14a}, we also find that satellites that are closer to their host galaxy at $z = 0$ tend to reionize earlier. This trend is not directly due to proximity to the host halo during reionization, as the relation is weaker (if present at all) at that time. Most satellites are reionized at the same time as the LG as a whole, during the global $x_{\rm m} \approx 0.4 - 0.6$.

Between 3 and 19 per cent of our present-day LG satellites show SF prior to the end of reionization, depending on the source model. The MW and M31 have similar numbers of these satellites with ancient populations, though M31 has 20 per cent more present-day satellites. Given that a significant fraction of the satellites \emph{observed} today exhibit ongoing SF both prior to and after reionization, our model with only high-mass ($>\!10^9\,\msun$) capable of SF is disfavoured. Since we do not have SF between the EoR and $z = 0$, we cannot compare full SFHs with observational data.

In general, a weak relation between the present-day distance from the central galaxy and cSF is present, i.e. nearer satellites have more cSF prior to the end of reionization. This trend is more pronounced at later times, meaning the proximity to the host galaxy progenitors at the time of reionization is a weaker predictor of cSF than the present-day distance. This correlation also indicates that environmental effects play a large role in observable quantities, emphasizing that constrained simulations are the best option for simulating reionization in the LG. Most significantly, the cSF prior to the end of reionization is not directly related to the mass and exhibits substantial scatter even at similar masses.

We find some differences between the MW and M31 cSF for the their satellites, though we emphasize that the distinction between the two is somewhat arbitrary. The mass range for the present-day satellites is larger for the MW than M31. We also find that the MW has a slightly higher fraction of satellites with ancient stellar populations. The cSF versus the present-day distance to the host galaxy relation is, unexpectedly, tighter for M31 versus the MW. We do caution that we do not constrain these small scales to exactly reproduce the observed population of satellites today. These results are merely suggestive of potential differences between two similarly sized, paired haloes.

Unsurprisingly, many factors affect satellite properties. We demonstrate that a range reasonable assumptions for the effect of radiative feedback on low-mass haloes can lead to very distinct SFHs in satellites galaxies. We also show that the environment of a galaxy can affect its reionization history. This result emphasizes the need for constrained simulations in interpreting locally observed properties.

\section{Acknowledgements}
We thank Hannes Jensen for sharing \textsc{\small c2raytools} and Garrelt Mellema for \textsc{\small C$^2$-Ray} assistance and helpful comments. We gratefully acknowledge PRACE for awarding us computational time under PRACE4LOFAR grants 2012061089, 2014102339, and 2013091910, including access to resource Curie based in France at CEA. We also acknowledge the Gauss Centre for Supercomputing e.V. for funding this project by providing computing time on the GCS Supercomputer SuperMUC at Leibniz Supercomputing Centre with PRACE and under grant h009za. Some of the numerical computations were done on the Apollo cluster at The University of Sussex. This work was supported by the Science and Technology Facilities Council [grant numbers ST/F002858/1 and ST/I000976/1] and the Southeast Physics Network (SEPNet). AK and GY were supported by the {\it Ministerio de Econom\'ia y Competitividad} and the {\it Fondo Europeo de Desarrollo Regional} (MINECO/FEDER, UE) in Spain through grants AYA2012-31101 and AYA2015-63810-P. AK thanks the Consolider-Ingenio 2010 Programme of the {\it Spanish Ministerio de Ciencia e Innovaci\'on} (MICINN) under grant MultiDark CSD2009-00064. He also acknowledges support from the {\it Australian Research Council} (ARC) grant DP140100198. YH was supported by the Israel Science Foundation (1013/12).

\bibliographystyle{mnras} 
\bibliography{../science_bib}

\end{document}